\documentclass{aa}
\newcommand{\be} {\begin{equation}}

\newcommand{\ee} {\end{equation}}
\newcommand{\etal}{{\it et al. }}
\newcommand{\src}{1WGA\,J1958.2+3232}

\newcommand{\bc}{\begin{center}}
\newcommand{\ec}{\end{center}}
\newcommand{\rc}{\rm}
\input psfig.tex
\begin{document}

\thesaurus{08.02.3; 08.05.2; 08.14.1; 08.16.7 \src; 13.25.5}

\title{The identification of the long--period X--ray pulsar 
\src\  with a Be--star/X--ray binary\,\thanks{Based on data collected  
at the Astronomical Observatory of Loiano, Italy}}

\author{G.L. Israel\inst{1, }\thanks{Affiliated to I.C.R.A.} \and 
S. Covino\inst{2} \and V.F. Polcaro\inst{3} \and L. Stella\inst{1,\star\star} 
}

\institute{Osservatorio Astronomico di Roma, Via Frascati 33, 
I--00040 Monteporzio Catone (Roma), Italy
\and
Osservatorio Astronomico di Brera, Via E. Bianchi 46, 
I--23807  Merate (Lecco), Italy
\and
Istituto di Astrofisica Spaziale, Area di Ricerca di Roma--Tor Vergata, 
Via Fosso del Cavaliere, I--00133 Roma, Italy
}

\date{Received 24 February 1999 / Accepted 11 March 1999 }
\offprints{G.L. Israel: gianluca@coma.mporzio .astro.it}
\maketitle
\markboth{Israel \etal; \src: a new Be--star/X--ray binary}
         {Israel \etal; \src: a new Be--star/X--ray binary}

\begin{abstract}
We present the results of optical observations performed between May 
and September 1998 of the stars within the position error circle 
of the recently discovered $\sim$12\,min pulsating X--ray source \src. 
Based on  photometry and slitless spectroscopy, we 
selected a likely optical counterpart, which was subsequently determined to 
be a m$_V$=15.7 B0Ve star, the spectral properties of which are in 
good agreement with the X--ray results. 
The proposed optical counterpart shows several H, He and Fe 
emission--lines, while the interstellar absorption spectral features 
place the star at a distance of $\sim$ 800 pc.
The inferred X--ray luminosity for this distance is $\simeq$ 10$^{33}$ 
erg s$^{-1}$ in the 2--10 keV band.
We conclude that \src\ is a {\rc likely} new long--period low--luminosity 
accreting neutron star in a Be--star/X--ray binary system.  
 
\end{abstract}


\section{Introduction}

The X--ray source \src\ was serendipitously detected on May 1993 within  the 
field of view of the Position Sensitive Proportional Counter 
(PSPC; 0.1--2.4 keV) in the focal plane of the ROSAT X--ray telescope. 
Highly significant pulsations at a period of 721$\pm$14~s were   
discovered in the ROSAT data (Israel \etal 1998). An ASCA observation 
performed on May 1998 detected \src\ at the flux level expected  
from the ROSAT pointing and confirmed the presence of a strong 
periodic signal at 734$\pm$1~s (Israel \etal 1999). 
A luminosity of $\sim$ 10$^{33}$($d$/1 kpc)$^2$ erg s$^{-1}$ in the 2--10 keV 
energy band was obtained (assuming an absorbed power--law model). Due to the 
large uncertainty in the period determined by ROSAT, it was not possible to 
determine whether the system contains an accreting magnetic white dwarf or a 
neutron star, based on the period derivative. Even the spectral characteristics 
were consistent with both scenarios. 
Accreting neutron stars in binary systems are often associated  
with O--B stars, while cataclysmic variables with K--M main sequence 
companion stars; in both cases strong emission--lines are expected to 
be detected. {\rc So far no unambiguous association of an accreting white dwarf 
to an OB star has been found. Expected X--ray luminosities are in 
the $\leq$10$^{32}$ erg s$^{-1}$ range for wind accretors}. Identifying the 
optical counterpart of \src\ and studying its spectrum 
provides decisive clues on the nature of system. 
 
We present here the results of an optical program aimed at studying the stars 
included in the X--ray 30\arcsec\ radius error circle of \src.  
The observations were performed between May and September 1998 at the Loiano 
Astronomical Observatory. In order to select objects with peculiar emission--lines, 
as expected from the companion star of this kind of binary systems,  
slitless multiobject spectroscopy as described by Polcaro \& Viotti (1998) 
was used.
This method allows to obtain a good spectrum for a large number of stars
and quickly select stars with strong emission--lines, down to magnitudes of m$_V$ $\leq$ 
16--18. 
Moreover the absence of a slit eliminates the light loss due to poor seeing,  
while  sky and nebular lines are spread out over the whole image, resulting only 
in a small increase of the background level.  

A Be spectral--type star was found well within the X--ray error circle. {\rc 
The probability of finding by chance a 
Be star with V$\leq$16.0\,mag within the small position 
uncertainty region is $\sim$10$^{-6}$}. Therefore the Be star represents a very 
likely optical counterpart of \src, making this source one of the few accreting 
X--ray pulsars with a pulse period P $>$ 500\,s in a Be/X--ray binary system.  

\section{Observations and results}

The observations were all performed with the 1.52\,m ``Cassini" telescope equipped 
with the Bologna Faint Objects Spectrometer and Camera 
{\it BFOSC} (Bregoli \etal \cite{BFMFO87}, Merighi \etal \cite{MMCMBO94}). 
During the first run (May 1998) the camera was equipped with the Thomson 
$1024\times1024$ CCD 
with 0.56\arcsec\ pixel size and $\sim 9\arcmin.6 \times 9\arcmin.6$ field of view 
(FOV). On July and September 1998 a Loral $2048\times2048$ CCD detector
with 0.41\arcsec\  pixel size and a FOV of $\sim 13\arcmin .5 \times 13\arcmin .5$ 
was used instead. We performed V, R and I photometry,  and low--resolution 
spectroscopy. The data were reduced using standard {\it ESO--MIDAS} and 
{\it IRAF} procedures for bias subtraction, flat--field correction, and one 
dimension stellar and sky spectra extraction. Cosmic rays were removed from 
each image and spectrum and the sky--subtracted stellar spectra were obtained, 
corrected for atmospheric extinction and flux calibrated (when possible). 
\begin{figure}
\centerline{\psfig{figure=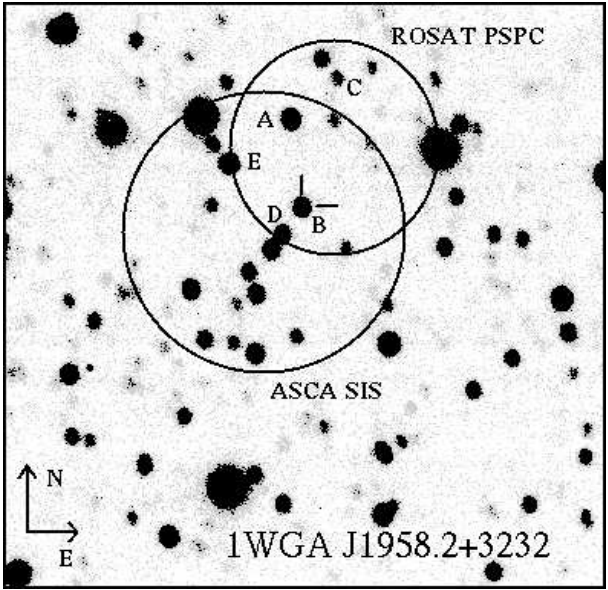,width=8cm,height=7.8cm} }
\label{fig:Robs}
\caption{2\,min image (filter R) of the field of 1WGA\,J1958.2+3232. The $H_\alpha$ 
emitter position is marked (Cand B). The position uncertainty circles of ROSAT PSPC 
(30\arcsec) and ASCA SIS (40\arcsec) are also shown.}
\end{figure}

\subsection{Photometry and Spectroscopy}

V, R, and I images of the whole $9\arcmin.6\times 9\arcmin.6$ wide 
1WGA\,J1958.2+3232 field were first obtained on 1998 May 30. In Figure\,1 the 
ROSAT PSPC error circle is shown together with the 
ASCA error circle (Israel \etal 1998, 1999). Note that at the time of the first optical 
observation only the ROSAT PSPC position uncertainty was known.
Photometry for each stellar object in the image was derived 
with the DAOPHOT\,II package (Stetson \cite{S87}),  
and Color-Magnitude Diagrams (CMDs) were then computed for all color
combinations. By the analysis of the $V$-$R$ vs. $V$ CMD 
a well defined main sequence is clearly visible (Fig.\,2), as expected in an area 
projected along the galactic plane like the one of \src\ ($l^{II} \sim 69^\circ$; 
$b^{II} \sim 2^\circ$). However, a number of objects with peculiar colors 
(redder or bluer) with respect the bulk of the stars were selected; 
three of them (stars A, B and C) were found to lie within the 
ROSAT position uncertainty region (see Table 1).  
\begin{table}
\begin{center}
\caption{Position and V magnitude of the studied stars within the uncertainty 
region.}
\begin{tabular}{cccc}
\hline
\hline
Star & R.A. (2000)$^a$ & DEC (2000)$^a$ & V$^b$ \\
&       (hh mm ss)  &  ($\deg$ $\arcmin$ $\arcsec$) & mag  \\
\hline
A & 19 58 14.7 & +32 33 07 & 15.4 \\
B & 19 58 14.4 & +32 32 42 & 15.7 \\
C & 19 58 13.6 & +32 33 16 & 17.0 \\
D & 19 58 16.0 & +32 32 54 & 16.6 \\
E & 19 58 14.9 & +32 32 34 & 16.0 \\
\hline
\end{tabular}
\end{center}
\noindent $^a$ calibrated with DSS1 plates. Uncertainty 1\arcsec. \\
\noindent $^b$ calibrated with DSS1 plates. Uncertainty 0.1 mag. 
\end{table}
Comparison between photometry obtained at the beginning and the end
of the run showed no signs of variability to a limit of $\sim 0.2$ mag for any
of the selected objects. A comparison between May and September 1999 
gave similar results. 

On 1998 May 30 we obtained {\rc a} low--resolution (20 \AA) slitless spectroscopic image 
of the field with an R filter and a large band 
grism covering the spectral region (4000--8000 \AA). We note that, in general, 
the choice of the grism, filter and time exposure depends 
on the grism dispersion and CCD size (which set the maximum 
spectral range allowed), and the crowding level of the field (which 
sets the maximum number of non--overlapping spectra to be analysed). The selected 
filter and grism combination gives a bandpass of $\sim$ 1000 \AA, centered on 
H${\alpha}$. 
A detailed analysis of the spectra obtained in this way allowed us to single 
out a relatively strong H$\alpha$ emission line associated to one (star B) of 
the three stars previously selected.

Slit spectroscopy was performed over selected stars on 1998 
May 30 -- June 2, July 27 -- 31, and September 14 -- 15 (see Table 2). 

\begin{table}
\begin{center}
\caption{Journal of the slit spectroscopic observations.}
\begin{tabular}{cccc}
\hline
\hline
 Date & UT & Wavelenght & Dispersion  \\
  &      & Range (\AA) &  (\AA/pixel) \\
\hline
1998 May 30 & 23:40    & 3900--7900 & 3.8     \\
"           & 00:30    &  "   & "\\
"           & 01:00    &  "   & "\\
"           & 01:30    &  "   & "\\
1998 May 31 & 23:30    &  "   & "\\
"           & 00:15    &  "   & "\\
1998 June 01& 00:10    & 6200--7900 & 1.7\\
"           & 01:00    & 3900--7900 & 3.8\\
1998 July 27& 23:11    & 3800--9000 & 2.5 \\
1998 July 28& 00:09    &  "   & " \\
1998 July 30& 20:25    &  "   & " \\
1998 July 31& 22:56    &  "   & " \\
1998 Sep. 14& 20:32    &  "   & " \\
1998 Sep. 15& 18:38    &  "   & "\\
\hline
\end{tabular}
\end{center}
\label{tab:obslog}
\end{table}

\subsection{Results}

The spectra of star A  are undoubtfully those of a classical
OB star. However neither emission--lines or other peculiarities are present that
would associate this star {\rc with} the X--rays source. 
Similar results were obtained for stars D and E for which a slitless spectrum was 
obtained on May 1998.

Due to its faintness (R=17.0) star C is more difficult to study. Even after a  
1.5\,h exposure spectrum, the S/N ratio of the spectrum barely reached the unity 
with our instrumental set--up. 
The steep rise of its
spectrum in the UV argues for a very hot object. However, it is unlikely that
it is to be associated with the X--ray source since no obvious 
emission features are present. Moreover, after the ASCA observation, its position 
lies outside the {\rc intersection region} of the two X--ray error circles.
\begin{figure}[tbh]
\centerline{\psfig{figure=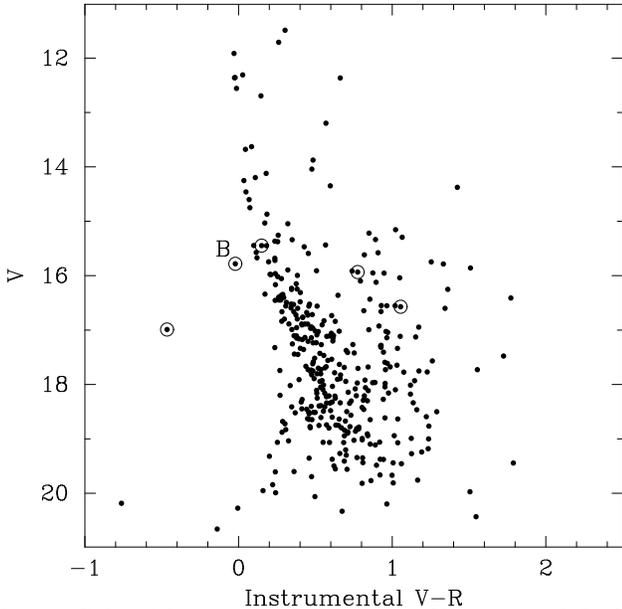,width=8cm,height=8.cm}}
\caption{Color--Magnitude diagram in filter V and R for \src. Big circles mark 
the five stars within the X--ray position uncertainty region, while the letter B 
represents the likely optical counterpart.}
\end{figure}

We note that at the border of the ROSAT error circle lies a bright (V$\simeq$12) star.
Its 4000--8000 \AA\ spectrum ($\Delta$$\lambda$=5.5\AA) was taken
in an earlier observation (June 1996) {\rc also} from the Loiano Observatory using the 
{\it BFOSC}; the star resulted to be a strongly reddened B8V spectral--type star, 
without any spectral peculiarity. 

The spectrum of star B clearly shows a very 
high ionization state (see Fig.\,3). The Balmer series lines 
are all in emission, up to the blue edge of our spectrogram as well as many 
He I ($\lambda$$\lambda$ ~4471, 4922, 5875, 6678, 7065, 8361 \AA) and  He II 
($\lambda$$\lambda$ 4686, 5412 \AA). 
An emission feature centered at 4634 \AA\ is present, that could be attributed
to  the N III 4634--40 \AA\ doublet, but we cannot exclude the possibility that
these feature is mainly due to Fe II lines. The presence of He II 
emission rules out a classical Be star, whereas  
the presence of emissions up to H7 exclude that of an Of star 
(see e.g. Jaschek \& Jaschek, 1987).
\begin{figure*}[tbh]
\centerline{\psfig{figure=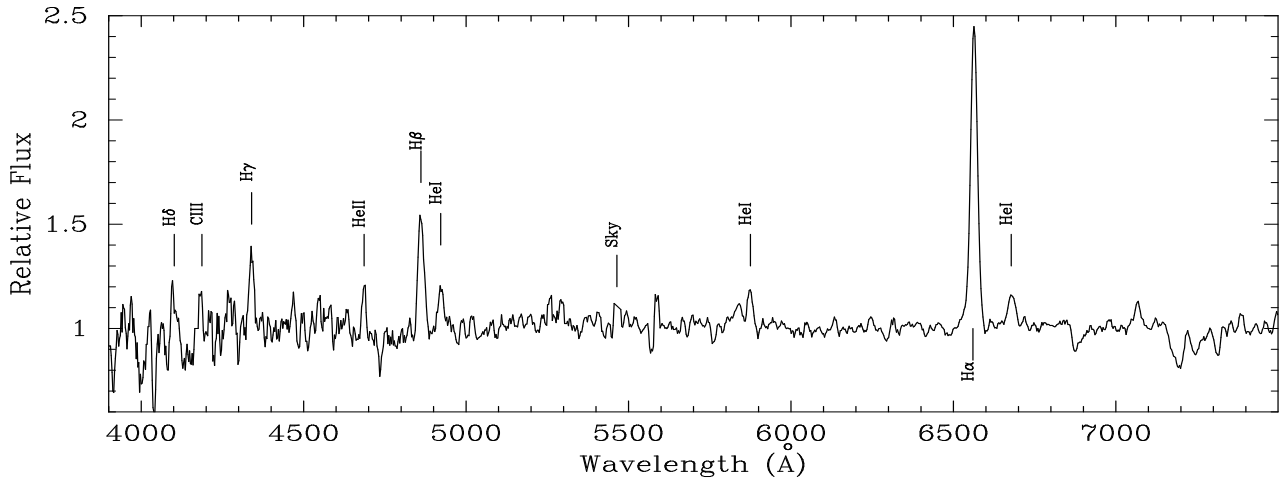,width=11cm,height=7.cm}}
\caption{5 hr low--resolution normalised spectrum of \src\ {\rc obtained} on July 
1998. The strongest emission lines detected in the spectrum are marked.}
\end{figure*}

Table\,3 gives the parameters of the strongest emission lines. 
These  lines (H$\alpha$, H$\beta$, H$\gamma$, He I 5875 \AA, He II 4686 \AA) 
show extended and asymmetric profiles, with a red wing more pronounced 
than the blue wing. At our spectral resolution, this might indicate 
a line splitting, possibly due to the presence of a disk. 
The absence of forbidden lines rules out the possibility of a pre--main 
sequence object or a cataclysmic variable. All this evidence clearly point to 
the association of this object with the X--ray source. 
\begin{table}
\begin{center}
\caption{List of the strongest spectral lines identified in the likely 
optical counterpart of the \src\ star.}
\begin{tabular}{ccc}
\hline
\hline
Line ID & Wavelength (\AA) & EW (\AA) \\
\hline
H$_{\delta}$ & 4101 & 1.6 \\
H$_{\gamma}$ & 4340 & 5.3 \\
H$_{\beta}$  & 4861 & 14.4\\
H$_{\alpha}$ & 6563 & 45.2 \\
HeI  & 4471 & 2.2 \\
HeII & 4686 & 2.5 \\
HeI  & 4922 & 3.7 \\ 
HeI  & 5876 & 2.5 \\
HeI  & 6678 & 6.0 \\
HeI  & 7066 & 6.2 \\
FeII & 4990 & 2.2 \\
FeII & 5466 & 8.0 \\
CrI  & 5297--8 & 2.5\\
\hline
\end{tabular}
\end{center}
\end{table}

\section{Discussion}

The spectrum of star B is similar to those of massive Be/X--ray binaries. 
As usual, a more detailed spectral classification is made more difficult by the 
fact that most of the classical criteria are unusable, since 
the H and most of the He lines are in emission and the Fe II complex masks 
most of the stellar atmospheric features. However, the presence of N III, 
C III and O II lines, and the $\sim$ 0.1 \AA\ equivalent width 
of {\rc the} Mg II 4481 \AA\ line suggest a B0 spectral type (see e.g. Jaschek 
\& Jaschek, 1987).
The few absorption lines that are clearly visible are wide, thus suggesting a
main sequence star. We can thus conclude that most probably the optical
companion of the collapsed object is a B0V star.

The equivalent width $\simeq$0.5 \AA\ of the interstellar Na II (5890 \AA) 
indicates an intermediate 
reddening ({\it E$_{B-V}$} $\simeq$ 0.6; following Hobbs, 1974), corresponding,
in the Cygnus region to a distance of $\sim$ 800 pc (Ishida, 1969). 
We also note that a {\it E$_{B-V}$} $\simeq$ 0.6 translates into a 
hydrogen column of $\sim$ 3 $\times$ 10$^{21}$ cm$^{-2}$ (Zombeck 1990), 
which is in good agreement with the N$_H$ values inferred from the 
spectral analysis of the merged ROSAT and ASCA data (N$_H$$\sim$4 $\times$ 10$^{21}$ 
cm$^{-2}$; Israel \etal 1999). This result implies a 1--10 keV X--ray luminosity of 
1.2 $\times$ 10$^{33}$ erg s$^{-1}$. {\rc Even though the optical data argue against 
an accreting white dwarf, only a measurement of the spin period 
derivative will firmly asses the nature of the accreting object. 
If the system hosts an accreting neutron star, than it would be}
one of the faintest Be/X--ray pulsars. 
Its closest analogoue {\rc would be} X Per which shows a 1--10 keV X--ray 
luminosity of 0.7--3.0 $\times$ 10$^{34}$ erg s$^{-1}$ in its low  
state (Haberl \etal 1998). Two recently identified Be/X--ray pulsars, namely
RX\,J0440.9+4431/BSD\,24--491 and RX\,J1037.5--564/LS\,1698 (Reig \& Roche, 1999),  
also share similar optical and X--ray characteristics with \src.  
{\rc If the accreting object is a white dwarf, \src\ would be the first example of 
a Be/white dwarf interacting binary system}.

\section{Conclusion}

Based on the optical results we obtained for the stars within the error circle of 
\src, we conclude that star B, a B0Ve spectral--type star, represents the very 
likely optical counterpart of \src. We compared the optical findings with  
recent ASCA X--ray data obtained for this source (Israel \etal 1999), other similar 
X--ray sources (Haberl \etal 1998; Reig \& Roche, 1999), {\rc and the current knowledge 
on accreting white dwarfs. We conclude that \src\ is a new likely faint}   
Be/X--ray pulsar, probably belonging to a new subclass of Be X--ray binary systems 
with long periods, persistent, low luminosities X--ray emission, and small flux 
variations. {\rc If the accreting white dwarf interpretation proves instead correct, 
than \src\ would be the first unambiguous example of a Be/white dwarf binary system.}  

\begin{acknowledgements} 
The authors thank R. Gualandi and S. Bernabei for their help during 
observations. The authors also thank F. Haberl the comment of which helped to 
improve this paper. 
This work was partially supported through ASI grants. 
\end{acknowledgements}


\begin{thebibliography}{}
\bibitem[1987]{BFMFO87} Bregoli, G., Federici, L., Merighi, R., \etal, 1987, in: ESO--OHP 
Workshop on the Optimization of the Use of CCD Detectors in Astronomy, 
Saint--Michel--l'Observatoire, France, June 17--19, 1986, Proceedings (A88--13301 
03--89). Garching, ESO, Germany, p. 177
\bibitem{} Haberl, F., Angelini, L., Motch, C. \etal 1998, A\&A, 330, 189
\bibitem{} Hobbs, L.M., 1974, ApJ, 191, 381 
\bibitem{} Ishida, K., 1969, MNRAS, 144, 55
\bibitem{} Israel, G.L., Angelini, L, Campana, S., \etal, 1998, MNRAS, 298, 502
\bibitem{} Israel, G.L., \etal, 1999, in preparation
\bibitem{} Jaschek, C.,  Jaschek, M., 1987, ``The Classification of Stars'', Cambridge
University Press 
\bibitem[1994]{MMCMBO94} Merighi, R., Mignoli, M., Ciattaglia, C., \etal, 
        1994, Bologna Technical Reports 09,-1994,-05
\bibitem{} Polcaro, V.F., Viotti, R., 1998, in: ``Astronomical Data Analysis Software 
and Systems VII'', ASP Conference Series, Vol. 145, Eds. R. Albrecht, R. N. Hook \& 
H. A. Bushouse
\bibitem { } Reig, P., Roche, P., 1999, MNRAS, in press (astro--ph/9902221)
\bibitem[1987]{S87} Stetson P.B. 1987, PASP, 99, 191
\bibitem{} Zombeck, M.V., 1990, in ``Handbook of space astronomy \& astrophysics'', Cambridge 
University Press
\end{thebibliography}
\end{document}